\documentclass[usenatbib,useAMS]{mnras}

\usepackage{mathptmx}
\usepackage{amssymb}
\usepackage[T1]{fontenc}

\DeclareRobustCommand{\VAN}[3]{#2}
\let\VANthebibliography\thebibliography
\def\thebibliography{\DeclareRobustCommand{\VAN}[3]{##3}\VANthebibliography}

\usepackage{fontawesome}
\usepackage{subfig}
\usepackage{ctable}
\usepackage{xspace}

\usepackage{graphicx}	
\usepackage{amsmath}	

\makeatletter
\newcommand{\github}[1]{%
   \href{#1}\textcolor{gray}{}{\faGithubSquare}%
}
\makeatother

\newcommand{\skirt}{{\sc skirt}\xspace}
\newcommand{\optuna}{\emph{Optuna}\xspace} 
\newcommand{\anngelina}{\emph{ANNgelina}\xspace}
\newcommand{\pytorch}{\emph{PyTorch}\xspace}

\title[Emulating MCRT with an ANN]{Emulating Radiative Transfer with Artificial Neural Networks}

\author[Snigdaa S.\,Sethuram et al.]{Snigdaa S.\,Sethuram,$^{1,2}$\thanks{E-mail: snigdaa.ram@gatech.edu}
Rachel K.\,Cochrane,$^{2}$
Christopher C.\,Hayward,$^{2}$
\newauthor Viviana Acquaviva,$^{2,3}$
Francisco Villaescusa-Navarro,$^{2,4}$
Gerg\"o Popping,$^{5}$
and John H.\,Wise$^{1}$
\vspace{0.2cm}\\
$^{1}$Center for Relativistic Astrophysics, Georgia Institute of Technology, North Avenue, Atlanta, GA 30332, USA\\
$^{2}$Center for Computational Astrophysics, Flatiron Institute, 162 5th Ave., New York, NY 10010, USA\\
$^{3}$Department of Physics, CUNY NYC College of Technology, 300 Jay Street, Brooklyn, NY 11201, USA\\
$^4$Department of Astrophysical Sciences, Princeton University, 4 Ivy Lane, Princeton, NJ 08544, USA\\
$^5$European Southern Observatory, Karl-Schwarzschild-Str. 2, 85748, Garching, Germany}

\date{Accepted XXX. Received YYY; in original form ZZZ}

\pubyear{2023}

\graphicspath{{./}{figures/}}

\begin{document}
\label{firstpage}
\pagerange{\pageref{firstpage}--\pageref{lastpage}}
\maketitle

\begin{abstract}
Forward-modeling observables from galaxy simulations enables direct comparisons between theory and observations. 
To generate synthetic spectral energy distributions (SEDs) that include dust absorption, re-emission, and scattering, Monte Carlo radiative transfer is often used in post-processing on a galaxy-by-galaxy basis.
However, this is computationally expensive, especially if one wants
to make predictions for suites of many cosmological simulations. To alleviate this computational burden, we have developed a radiative transfer emulator using an artificial neural network (ANN), \anngelina, that can reliably predict SEDs of simulated galaxies using a small number of integrated properties of the simulated galaxies: star formation rate, stellar and dust masses, and mass-weighted metallicities of all star particles and of only star particles with age $<10$ Myr.
Here, we present the methodology and quantify the accuracy of the predictions. We train the ANN on SEDs computed for galaxies from the \emph{IllustrisTNG} project's TNG50
cosmological magnetohydrodynamical simulation. \anngelina is able to predict the SEDs of TNG50 galaxies in the ultraviolet (UV) to millimetre regime with a typical median absolute error of $\sim 7$ per cent. The prediction error is the greatest in the UV, possibly due to the viewing-angle dependence being greatest in this wavelength regime.
Our results demonstrate that our ANN-based emulator is a promising computationally inexpensive alternative for forward-modeling galaxy SEDs from cosmological simulations.
\end{abstract}




\section{Introduction}\label{sec:intro}

Forward-modeling observables from galaxy formation simulations provides a means to confront theoretical models with observations directly.
This approach represents an alternative to the more-traditional method of inferring physical properties such as stellar mass and 
star formation rate (SFR) from observations and comparing those with the corresponding quantities from simulations.
One advantage of forward-modeling observables from a simulation is the avoidance of ``throwing away'' information from the simulation; e.g.\,when
predicting spectral energy distributions (SEDs), the full star formation history of a simulated galaxy is used.

\begin{figure}
    \centering
    \includegraphics[width=\linewidth]{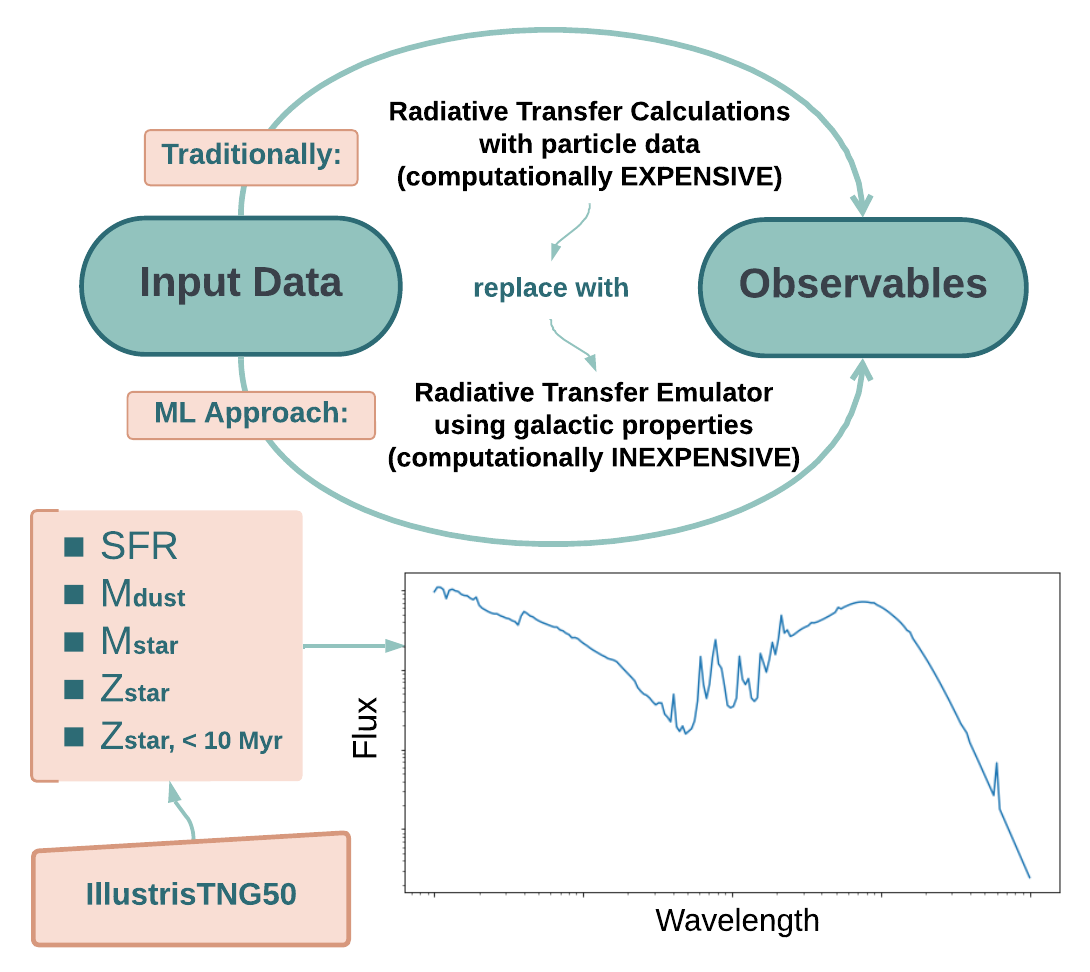}
    \caption{\anngelina project workflow. In the traditional approach, computationally expensive MCRT calculations are needed to forward-model SEDs and images from simulated galaxies. In this work, we demonstrate how an ANN can be used to predict SEDs of simulated
    galaxies, accelerating this process by at least 7 orders of magnitude.}
    \label{fig:workflowschem}
\end{figure}

One particularly common forward-modeling technique in the galaxy formation community is to predict ultraviolet-to-millimeter (UV-to-mm) SEDs of simulated galaxies,
including both integrated SEDs and images in various observed bands
(e.g.\,\citealt{Jonsson2006,Jonsson2010,skirt15}; see \citealt{stein13_rtreview} for a review).
This calculation is normally done by performing dust radiative transfer on simulated galaxies in post-processing to compute how light from star particles
propagates through the simulated galaxy's interstellar medium (ISM) and is absorbed, scattered, and re-emitted by interstellar dust.
This approach has been applied to compare model predictions with observations of various classes of objects such as massive galaxies or active galactic nuclei \citep[e.g.][]{Hayward2012,Lanz2014,
Narayanan2015,Safarzadeh2017irx,Cochrane2019,Cochrane2022,Cochrane2023,Cochrane2023b,Baes2020,Parsotan2021}.
Moreover, such synthetic observations can be used to perform ground-truth experiments to test and refine techniques for inferring
physical quantities from observations \citep[e.g.][]{Wuyts2010,HS2015,Michalowski2014,
SH2018,McKinney2021,Cochrane2022}.\\
\indent Unfortunately, this forward-modeling step can incur significant additional computational expense and requires detailed outputs from simulations
(the full 3D density distribution of stars and dust, for example). This required detailed information is not always available, in particular for
coarse-resolution simulations.
In such scenarios, the Monte Carlo radiative transfer (MCRT) calculations must make various assumptions (e.g.\, regarding sub-grid dust clumpiness)
that can affect the robustness of the predicted observables.
It is thus desirable to find a computationally inexpensive, robust method for making such predictions from a limited amount of data, such as integrated galaxy properties. Having a fast method available to run radiative transfer with varying input parameters would allow predictions to be made for different input parameters, which will ultimately allow us to marginalize over the uncertainties associated with these parameters.\\
\indent Several previous works \citep{hayw11,saf16,Lovell2021,Cochrane2023} have demonstrated that it is possible to predict observed-frame far-IR (FIR) fluxes of simulated galaxies
using only a small number of integrated properties of the galaxies, such as SFR and dust mass.
The success of these approaches suggests that geometric variations amongst simulated galaxies
play a subdominant role in determining their FIR SEDs and gives confidence that one can employ such simple approaches
as an alternative to full MCRT, as has been done in a handful of works
\citep[e.g.][]{HB2013,Safarzadeh2017imf,Hayward2021,Miller2015,Popping2020,Cochrane2023}.
However, these works have focused on the thermal dust emission; it is likely more difficult to predict full UV-mm SEDs
because geometry plays a greater role in the UV-optical due to the non-isotropic nature of dust attenuation on galaxy scales.
Nevertheless, in this work, we aim to predict UV-mm SEDs using only a small number of integrated properties of simulated galaxies,
without specifying any information about the galaxy geometry. The general workflow of the project is shown in Figure \ref{fig:workflowschem}.


Using machine learning (ML) techniques to emulate computationally intensive calculations in simulations has shown substantial promise in astrophysics \citep{buchner19, kasim21, astrid22} as well as other computationally intensive STEM fields, e.g. climate modeling \citep{weber20}. 
Motivated by such works and the demonstrated possibility of predicting FIR SEDs from a small number of galaxy properties,
in this work, we develop an artificial neural network (ANN) emulator for MCRT calculations on simulated galaxies.
NNs are deep learning (DL) algorithms, a type of ML that employs multiple layers of computation to learn trends and correlations in data. NNs are robust at understanding complex relationships between input and output data due to their iterative calculations that optimise the data fit in pieces. ML algorithms take into account not only the nonlinear mapping of input to output data but also the distribution of properties across the full training set \citep{relupaper}.\\
\indent In this paper, we explore the possibility of an ANN reproducing nonlinear relationships between galaxy properties and galaxy SEDs. Our work addresses the following question: ``given a set of 3D MCRT calculations with fixed assumptions (regarding e.g.\,the stellar initial mass function, single-age stellar population SED templates, and dust model) performed on galaxies selected from a single cosmological simulation, how well can an ANN emulate the MCRT calculations to predict integrated UV-mm SEDs of the simulated galaxies?''
The inverse workflow has been attempted, including using ML to derive galaxy properties from a given SED \citep{gilda21} and using ML to derive
star formation histories for galaxies in the \emph{Illustris} and {\sc eagle} simulations \citep{sfhML}, but to the best of our knowledge,
using an NN to predict galaxy SEDs has not been previously attempted. We use SEDs from the TNG50 cosmological magnetohydrodynamical simulation \citep{tng501, tng502}, which is the highest-resolution simulation
in the \emph{IllustrisTNG} suite (TNG) \citep{tng501, tng502}, to train our ANN, \anngelina\footnote{\url{https://github.com/snigdaa/runANNgelina}}.

The structure of this paper is as follows. In Section \ref{sec:methods}, we discuss the methods used to create \anngelina and present an overview of the TNG50 dataset used for training. In Section \ref{sec:results}, we present our preliminary results, touching on some of the interpretation that went into analyzing \anngelina's performance.
We discuss applications and limitations of \anngelina, as well as improvements to be made in future work. We draw conclusions in Section \ref{sec:conclusions}.

\section{Data} \label{sec:data}
The version of \anngelina released with this paper (ANNgelina\_v1.0) was trained on synthetic SEDs obtained by running a radiative transfer code on galaxies in the TNG50 \citep{tng501, tng502} simulation. Here, we provide an overview of the TNG simulations, the sample of galaxies selected from TNG50, and the methods used to model their SEDs.


\begin{figure*}
\centering
{\includegraphics[trim={0cm 0.5cm 0cm 0cm}, clip, width=\linewidth]{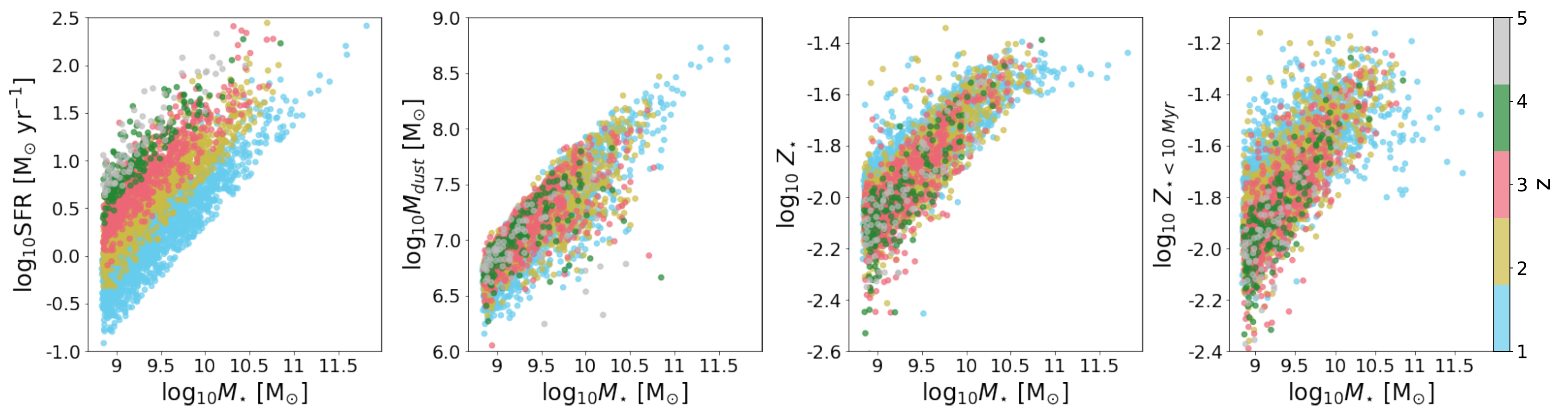}}\,
\vspace{-0.3 cm}  
\caption{Correlations between various integrated properties and stellar mass for all galaxies from TNG50 with $M_{\star} > 10^{9}$ M$_{\odot}$, illustrating the region of parameter space spanned by our simulated galaxy sample: SFR (\emph{left-most}), dust mass (\emph{middle-left}),
and mass-weighted metallicities of all stars (\emph{middle-right}) and young ($<10$ Myr-old) stars (\emph{right-most}).
In each panel, the points are color-coded by redshift.
All properties have been calculated within twice the stellar half-mass radius of each galaxy.
As expected, all of the considered quantities correlate with total stellar mass. The normalization of the SFR-$M_{\star}$ relation increases with increasing redshift, whereas the $M_{\rm dust}-M_{\star}$ correlation is independent of redshift for the range considered.
}
\label{fig:paramz}
\end{figure*}

\subsection{IllustrisTNG simulations} \label{sec:tng50sims}
The \emph{IllustrisTNG} project\footnote{\url{https://www.tng-project.org}} is a suite of large-volume cosmological magnetohydrodynamical simulations that model the formation of galaxies from
early times to $z=0$. The numerical methods and galaxy formation model used for the simulations are described in detail in \citet{pill18b}, \citet{wein17}, and \citet{pill18a}, so we will
only briefly summarize them here.

The simulations were run with {\sc arepo}\footnote{\url{https://gitlab.mpcdf.mpg.de/vrs/arepo}} \citep{Springel:2010arepo}, an unstructured moving-mesh hydrodynamics code.
Star formation is implemented by stochastically spawning stellar particles at a rate set by a volume density-dependent Kennicutt-Schmidt law \citep{Kennicutt:1998,Schmidt:1959};
a density threshold of 0.13 cm$^{-3}$ is employed. Stellar populations are evolved self-consistently, including chemical enrichment and gas recycling, resulting in mass loss.
An effective equation of state is used to approximate how SNe heat the ISM, and stellar feedback-driven winds are implemented by stochastically isotropically kicking
and temporarily hydrodynamically decoupling gas cells \citep{Springel:2003}. Black hole accretion is modelled as Eddington-limited modified Bondi-Hoyle accretion.
A two-mode AGN feedback model is employed: a relatively efficient kinetic channel at low Eddington ratio (`radio-mode') and a less-efficient
thermal channel at high Eddington ratio (`quasar mode') (see \citealt{wein17} for details).
The free parameters in the subgrid models employed were tuned to match the galaxy stellar mass function and stellar mass-to-halo mass relation at $z \sim 0$,
in addition to the cosmic star formation rate history at $z \la 10$. The black hole mass--stellar mass relation, halo gas fractions and stellar half-mass radii of galaxies
were also considered (see section 3.2 of \citealt{pill18b}).
The TNG simulations adopt a cosmology consistent with the Planck 2015 results \citep{planck16}: $\Omega_{m} = 0.31,\,\, \Omega_{\Lambda} = 0.69,\,\, \Omega_{b} = 0.0486,\,\,h = 0.677,\,\,\sigma_{8} = 0.8159$, and $n_{s} = 0.97$ \citep{tng501}.

\subsubsection{IllustrisTNG50 dataset} \label{sec:tng50}

We use the highest-resolution simulation from the \emph{IllustrisTNG} suite, TNG50 \citep{tng501}, which has a volume of $(35 \ h^{-1} \ {\rm Mpc})^3$.
The mass resolution is 
$8.5\times10^{4}\,\rm{M_{\odot}}$ for baryonic particles and $4.5\times10^{5}\,\rm{M_{\odot}}$ for dark matter particles.
TNG50 employs a collisionless softening length of $0.3\,\rm{kpc}$ at $z = 0$ and an adaptive gas softening length with a minimum of $74$ comoving pc.\\
\indent We draw our sample of TNG50 galaxies from \citet{popping}, who modelled integrated SEDs for approximately 3,800 star-forming galaxies\footnote{Here, a single halo is considered a distinct galaxy in each of the snapshots. Given the large redshift spacing between snapshots, this assumption is reasonable.} on
or above the star formation main sequence with $M_{\star}>10^{8.7}\,\rm{M_{\odot}}$. Our sample is comprised of $1709$ (central and satellite) galaxies at $z=1$, $1149$ galaxies at $z=2$, $620$ galaxies at $z=3$, $184$ galaxies at $z=4$, and $76$ galaxies at $z=5$. There are fewer SEDs available for the higher-redshift snapshots, since fewer galaxies meet the redshift-independent stellar mass criterion. Figure \ref{fig:paramz} shows the distribution of the parameter space covered by our TNG50 dataset. 

\begin{figure*}
    \centering
    {\includegraphics[trim={4cm 0 4cm 2cm},clip, width=\linewidth]{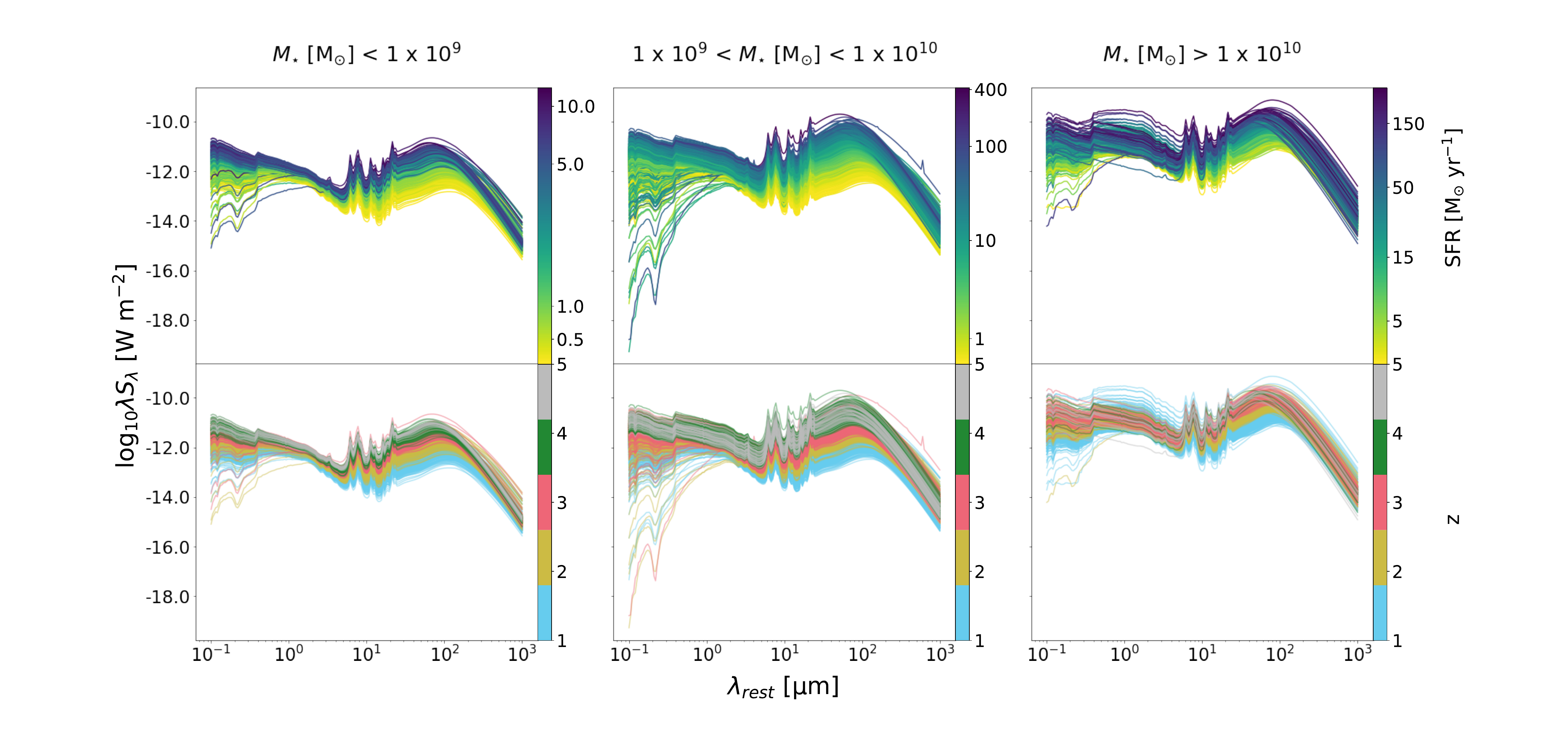}}
    \vspace{-0.8cm}
    \caption{ TNG50 SEDs used to
    train our ANN.
    The top row shows the SEDs color-coded by SFR, and the bottom row shows the SEDs color-coded by redshift; the columns show different stellar mass bins, as indicated at the tops of the figures.
    In a given mass bin, galaxies with higher SFRs tend to have higher IR fluxes, and their FIR SEDs are hotter. In the two lower-mass bins, the UV-optical fluxes also increase with SFR. This is less evident in the highest-mass bin
    due to increasing dust attenuation. In all mass bins, there is a subset of galaxies with high SFRs and heavily reddened UV-optical SEDs.
    Higher-redshift galaxies tend to have larger fluxes due to the evolution of the star formation main sequence (at fixed mass, SFR increases with increasing redshift).}
    \label{fig:binnedseds}
\end{figure*}

\subsection{Generation of synthetic SEDs}\label{sec:rttng50}
Synthetic galaxy SEDs were generated by \cite{popping} using the \skirt MCRT code \citep{skirt15}.\footnote{\url{https://skirt.ugent.be/root/_home.html}}
\skirt takes the three-dimensional dust and stellar density distributions from a galaxy simulation and propagates photon packets from radiation sources (i.e. star particles; AGN emission was not modeled)
through the simulated galaxies' ISM using a Monte Carlo approach to model dust absorption, scattering, and re-emission.
\citet{popping} followed the methods described by \citet{schulz20} to calculate the SEDs for TNG50 galaxies. We briefly summarise their methods here.\\
\indent Gas and stellar particles within $7.5$ times the stellar half-mass radius were extracted from subhalos.
Star particles with ages greater than 10 Myr were assigned \citet{bruz03}
single-age stellar population model template SEDs based on their age and metallicity.
Star particles with ages $\le10$ Myr were assigned SEDs from \citet{groves08}, which include a sub-grid model for HII and photodissociation regions. 
Dust was modelled using the gas-phase metal density distribution, with a constant dust-to-metals mass ratio of 0.4 \citep[e.g.][]{Dwek1998,James2002}. 
Gas cells with $T > 7.5 \times 10^4\,\rm{K}$ were assumed to contain no dust due to dust destruction via sputtering.
The \citet{weindraine} Milky Way dust model was employed, with a mix of graphite, silicate, and PAH grains.
Self-absorption of dust grains was not taken into account, but this is unlikely to be significant for the bulk of the galaxy population considered here
\citep[e.g.][]{hayw11}.\\
\indent The modeled SEDs span rest-frame $0.1-1000\,\micron$ and are sampled at 200 uniformly log-spaced wavelengths.
The simulated galaxies were ``observed'' ``face-on'' (i.e.\,the detector was placed such that the line of sight was along the angular momentum vector
of the simulated galaxy, corresponding to face-on projections for simulated disc galaxies).

\subsubsection{Modeled SEDs for TNG50 galaxies}\label{sec:rttng50_trends}
In Figure \ref{fig:binnedseds}, we show examples of modeled SEDs for TNG50 galaxies in several stellar mass bins. The upper panel shows SEDs color-coded by SFR. The lower panel shows SEDs color-coded by redshift. We use this visualization to understand trends in our data. As expected, higher SFRs generally yield higher fluxes. There are interesting trends at short wavelengths, though, with some high-SFR SEDs displaying particularly low fluxes at $\lambda \lesssim 1\,\micron$, owing to high levels of dust attenuation. We also note that SFR correlates with the shape of the FIR SED: the FIR SED peaks at shorter wavelengths for galaxies with higher SFRs. For a given stellar mass bin, higher-redshift (i.e. earlier-forming) galaxies are also brighter at all wavelengths (note that these SEDs are generated in the rest frame). This likely reflects the evolution of the ``main-sequence'': at a given stellar mass, higher-redshift star-forming galaxies have higher SFRs and hence higher intrinsic (pre-dust-attenuated) fluxes.

\section{Methods}\label{sec:methods}
We design an ANN, \anngelina, to emulate SED generation of TNG50 galaxies and hence bypass the radiative transfer procedure described in Section \ref{sec:rttng50}. An ANN architecture consists of an input layer with several ``features''; one or more hidden layers, in which the features are manipulated; and an output layer, which contains the desired outputs, or ``labels''. The structure of \anngelina is illustrated in Figure \ref{fig:anngelinaDiagram}. We use a fully connected network, where each neuron talks to all neurons in the previous and following layers. Each neurons in the input layer corresponds to a galaxy property, such as stellar mass or SFR, and each neuron in the output later corresponds to the flux at one wavelength. \anngelina is built with \pytorch \citep{pytorch}\footnote{\url{https://pytorch.org/}}. In the following sections, we describe the architecture and parameter choices in more detail and outline the training procedure.

\subsection{Data normalisation and partition}\label{sec:NN_architecture}
Each input feature, $X$, is normalised as follows:
\begin{equation}\label{eq:feature_norm}
X_{\rm{norm}} =  \frac{\log_{10}X - \mu(\log_{10}X)}{\sigma(\log_{10}X)}
\end{equation}
For any features that had zero values, we added a negligible nonzero value to that feature before normalizing to avoid taking the logarithm of zero. This normalization procedure accelerates the
training of the network.\\
\indent The flux is input as the irradiance, $E$, defined as  $E = \lambda S_{\lambda}/(\rm{W\,m^{-2}})$. We use $\log_{10}E$ as the labels in the network.
Data are then randomized and partitioned into training ($70\%$), validation ($15\%$), and test ($15\%$) sets.

\subsection{Training the ANN}\label{sec:NN_training}
Here, we provide a broad overview of the training process and various hyper-parameter and function choices. During a ``forward pass'', the ANN pushes the training data through the hidden layers by using the current values of the weights for each feature. After each forward pass, the model updates itself through a ``backpropagation'' phase, in which it determines a loss value (based on the difference between the predicted and true outputs) and updates the weights according to an optimization function, usually a type of gradient descent. The number of times the forward pass and backpropagation cycle is executed depends on the pre-defined batch size and number of epochs.\\
\indent Several parameters define the workflow of the network. Hyper-parameters such as the number of hidden layers and neurons in each layer, dropout fraction, weight decay, and activation function are related to the structure of the network. These hyper-parameters were optimized as described in Section \ref{sec:hyper_opt}. The learning rate, number of epochs, and batch size affect the training procedure; appropriate choices ensure generalizability of the model and improved efficiency while training. The batch size (128) and number of epochs (1500) were tuned manually. The batch size refers to the number of samples to be propagated through the network in one full pass. One epoch is completed once all training samples have been propagated through the network. The initial learning rate, dropout fraction, weight decay, number of hidden layers, and number of neurons per hidden layer were determined using \optuna \citep{optuna}, as described in Section \ref{sec:hyper_opt}.\\
\indent We use ``dropout'', a simple and efficient regularization technique that consists of removing a random subset of weights at every training epoch, thereby ensuring that the learned weights are more robust and preventing over-fitting. The dropout fraction determines what percentage of weights are removed. Weight decay works in tandem with dropout and is another regularization technique applied to all weights in a network. There is an extra term added to the loss function to reduce the variance in the network weights. \\
\indent We adopt the following optimizer, activation and loss functions:\vspace{0.2cm}\\
{\bf{Optimizer function:}} We chose the optimizer function AdamW, which is a modified version of the Adam optimizer \citep{adamopt} in which the learning rate and the weight decay of the model are optimized separately. It takes an initial learning rate and weight decay value and updates them for each feature's weight individually while performing gradient descent. It performs well at handling non-smooth data. The learning rate controls how much the model changes in response to predicted error during each forward pass.
\vspace{0.2cm}\\ 
{\bf{Activation function:}} We use the LeakyReLu (Leaky Rectified Linear Unit) activation function \citep{maas2013}, which is defined as
\begin{equation}
y = \begin{cases}
x&(x \geq 0), \\
0.01x&(x < 0).
\end{cases}
\end{equation}
\vspace{0.2cm}\\
{\bf{Loss function:}} NN losses in supervised learning models are determined by quantifying the difference between the predicted and true outputs using a loss function.
We adopt the mean square error (MSE) as our loss function. The MSE guarantees that the prediction represents the posterior mean without assuming any shape for the posterior distribution. During the training phase, the network minimizes the MSE averaged over each batch. The MSE for a batch of size $b$ is given by
\begin{equation}
\mathrm{MSE} = \frac{1}{200 \cdot b}\sum_{i=1}^{b} \sum_{\lambda = 1}^{200} \left[\log_{10}(E_{\lambda, \ i,\ \mathrm{pred}} / E_{\lambda,\ i,\ \mathrm{true}})\right]^2
\end{equation}
where $\lambda$ represents the wavelength index of the SED, given 200 wavelengths sampled between $0.1 -1000 \ \mu \mathrm{m}$, and $E_{\lambda,\ i}$ is the irradiance at a given wavelength, $\lambda$, for a given galaxy in the batch.

\begin{figure}
    \centering
    \includegraphics[width=\linewidth]{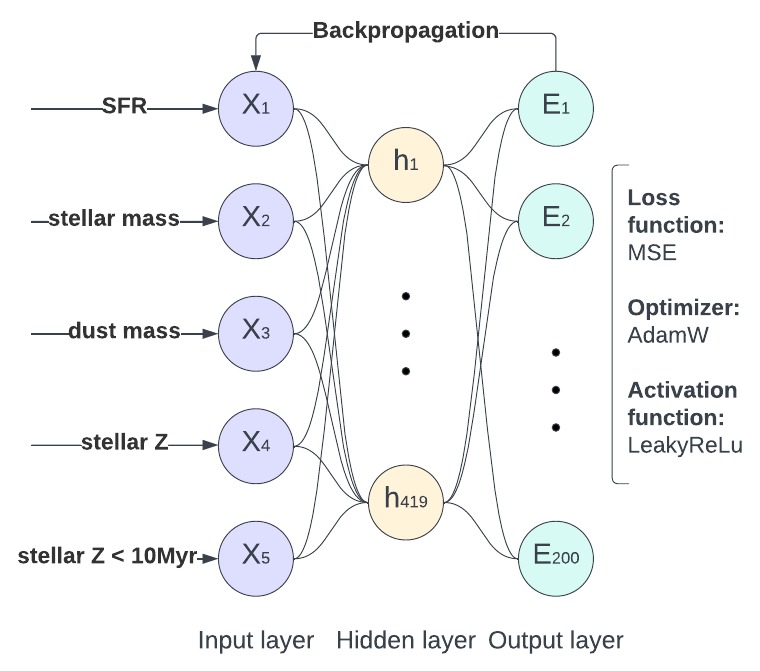}
    \caption{The architecture of \anngelina, with input features labeled $X_{1}-X_{5}$ (purple), neurons in the hidden layer labeled $\rm{h}_{1}-\rm{h}_{419}$ (orange), and output neurons labeled $\rm{E}_{1}-\rm{E}_{200}$ (green). Not all neurons are shown. The following galaxy properties are used as features: SFR, stellar mass, dust mass, and stellar metallicities of both all and young ($<10$ Myr old) stars. The output comprises rest-frame fluxes at $200$ wavelengths, uniformly spaced in log-space. We note the loss function, optimizer and activation functions used in training.}
    \label{fig:anngelinaDiagram}
\end{figure}

\ctable[
	caption = {Correlation matrix for TNG50 galaxy properties\label{tab:corrmatrixTNG}.},
	center, notespar, doinside=\small]{lcccccc}{}{
		\FL Feature & $M_{\rm dust}$ & $M_{\star}$ & $Z_{\star}$ & $Z_{\star <10 {\rm \,Myr}}$ & SFR  \ML
	    $M_{\rm dust}$ 		& 1.0 	& 0.88  & 0.41 	& 0.28 	& 0.48  \NN
		$M_{\star}$ 		& 0.88 	& 1.0 	& 0.44 	& 0.29 	& 0.48  \NN
		$Z_{\star}$ 		& 0.41 	& 0.44 	& 1.0 	& 0.9 	& 0.37  \NN
		$Z_{\star <10 \rm \,Myr}$ & 0.28 & 0.29 	& 0.9 	& 1.0 	& 0.25 \NN
		SFR & 0.48 	& 0.48 	& 0.37 	& 0.25 	& 1.0   \LL
}

\captionsetup[subfloat]{
justification=centering,
width=5.5cm
}

\begin{figure*}
     \centering
     \subfloat[][Among the best predictions in the test set, with an MSE corresponding to the $2^{\rm nd}$ percentile of the sample's distribution.]{\includegraphics[width=5.5 cm]{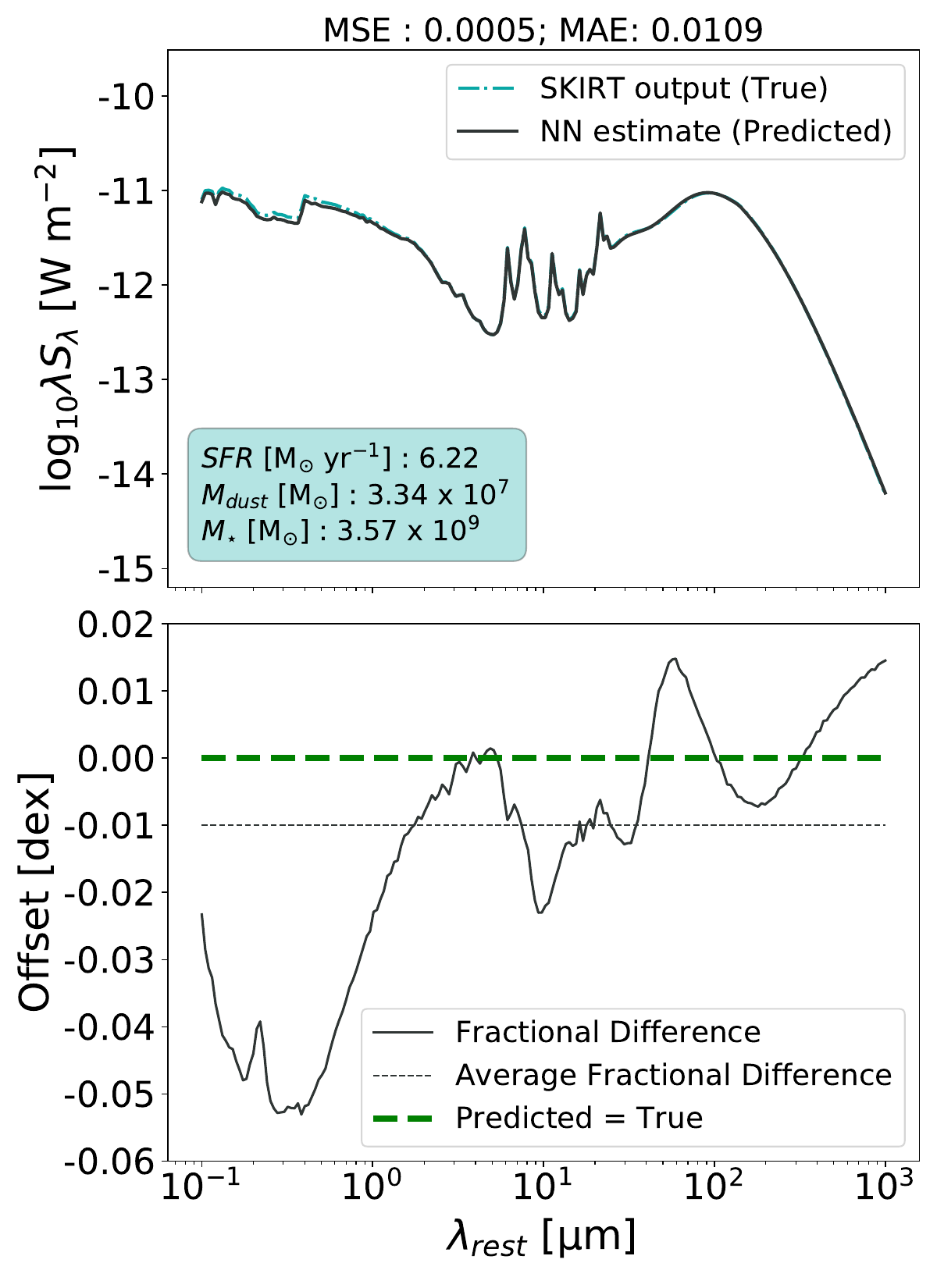}\label{fig:bestfit}}
     \subfloat[][A typical example ($50^{\rm th}$ percentile).]{\includegraphics[width=5.5 cm]{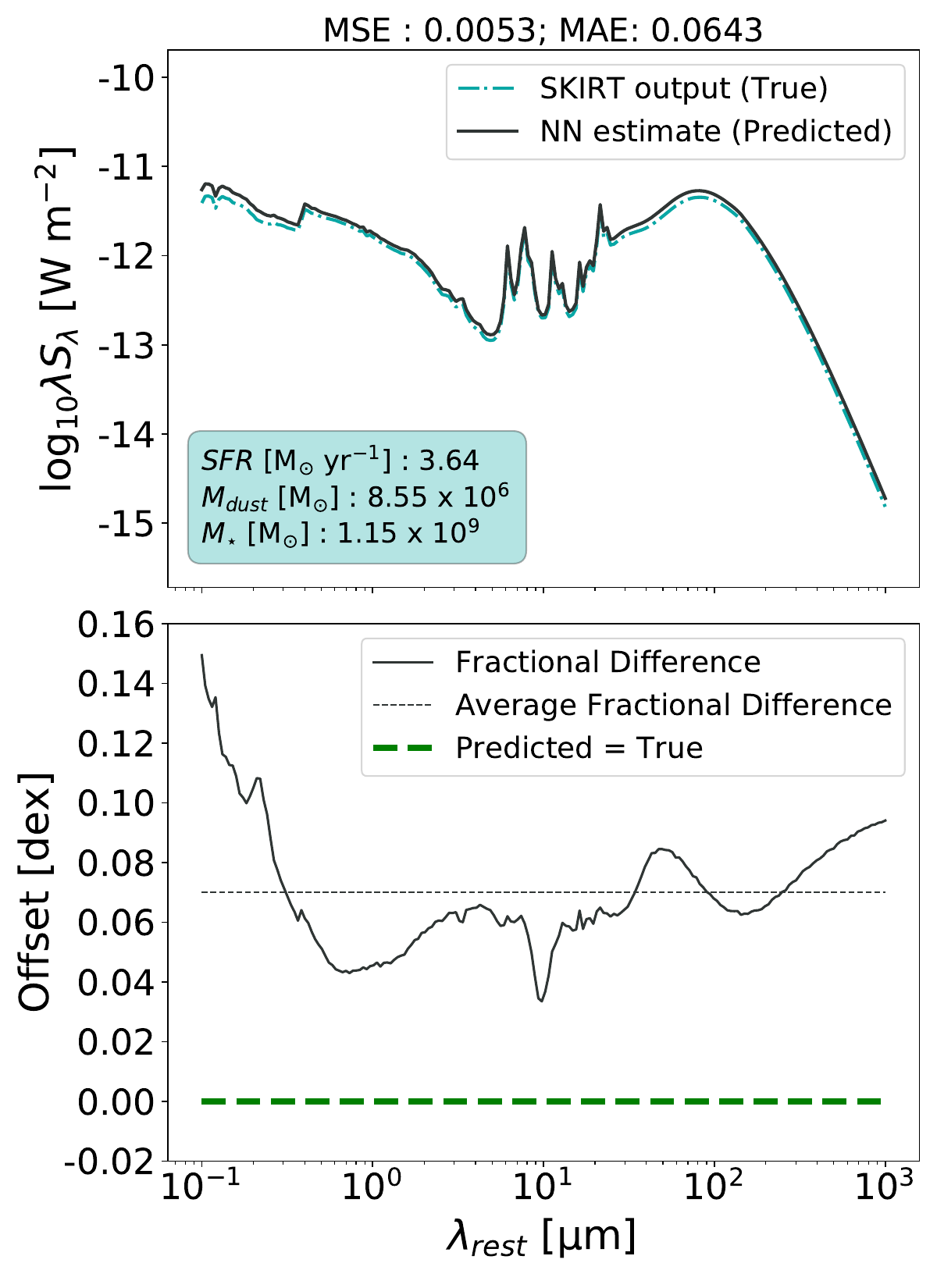}\label{fig:avgfit}}
     \subfloat[][One of the least successful predictions ($98^{\rm th}$ percentile).]{\includegraphics[width=5.5 cm]{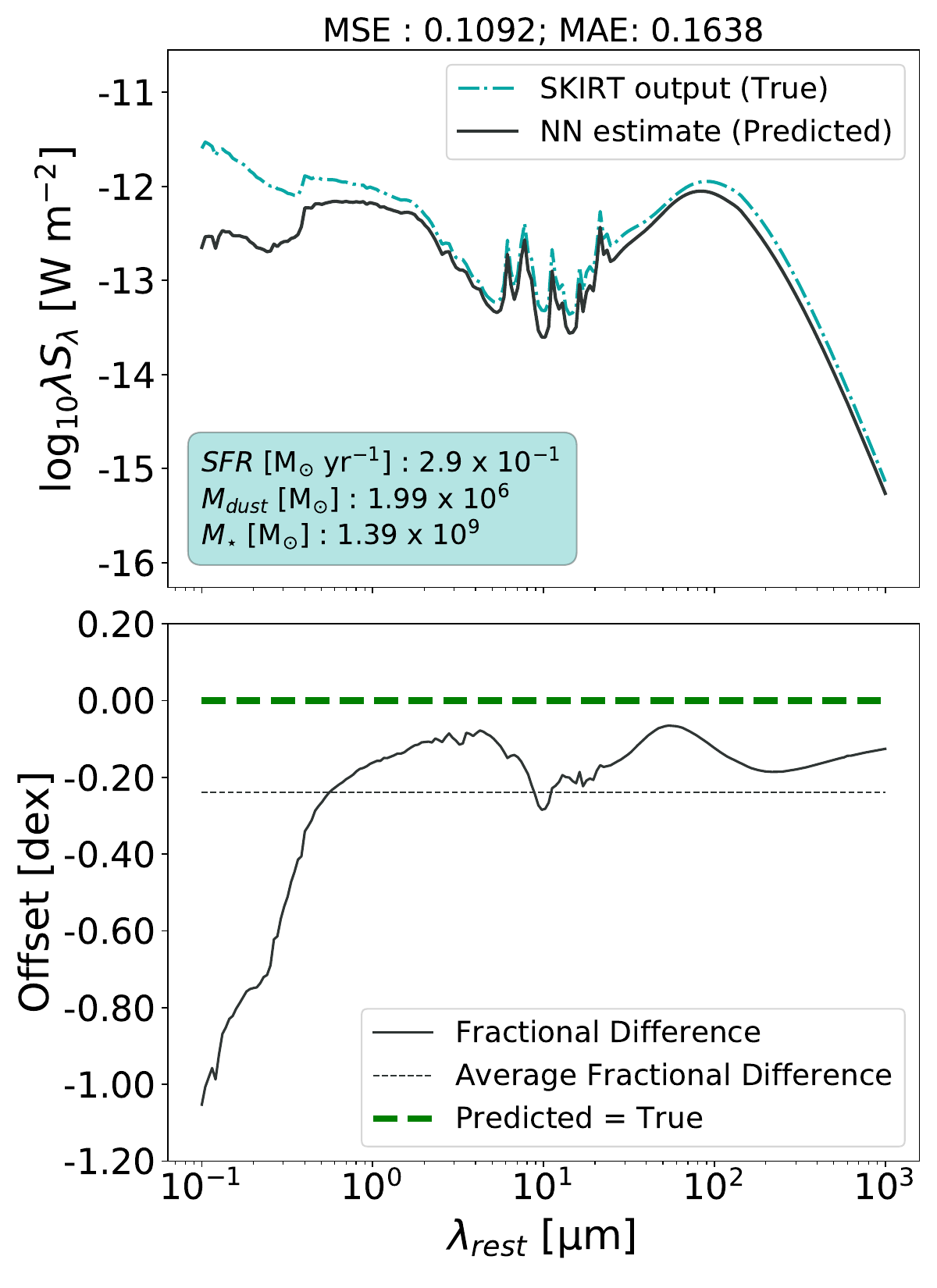}\label{fig:worstfit}}
     \caption{Three example SEDs predicted using \anngelina, with MSE increasing from left to right.
     In each column, the top panel shows the true SED computed using \skirt (cyan, dot-dashed) and that predicted using \anngelina (black).
     The bottom panels show the logarithm of the absolute fractional difference (black lines). The average value across all wavelengths is indicated via the horizontal short-dashed black lines; the green dashed horizontal lines indicate zero prediction error.
     The labels at the top of each column indicate the MSE and MAE (i.e.\,median absolute fractional difference) values.
     The typical absolute fractional difference between true and predicted SEDs is in the range from $0.03-0.07$. The average fractional difference lines indicate that our model has low bias, even for the worst example shown (right column).
     Note that the UV emission is more poorly predicted than any other part of the SED in all cases.}
\label{fig:nnfits}
\end{figure*}

\subsection{Hyperparameter optimization}\label{sec:hyper_opt}
We use the \optuna hyperparameter tuning package \citep{optuna} to tune the follwing model parameters: learning rate, dropout fraction, weight decay, number of hidden layers, and number of neurons per hidden layer. After testing with up to 11 hidden layers, each with up to 1000 neurons, optimal performance was achieved with a shallow, one hidden layer architecture with 419 neurons, as shown in Figure \ref{fig:anngelinaDiagram}. All hyperparameter values for our optimized model can be found in the public repository listed in Section \ref{sec:data}. 

\subsection{Galaxy feature selection tests}\label{sec:feature_selection}
This version of \anngelina uses just five galaxy properties (SFR, $M_{\rm dust}$, $M_{\star}$, $Z_{\star}$, and $Z_{\star < 10 \rm \,Myr}$) to predict SEDs. While designing \anngelina, we tested five additional  galaxy properties: gas mass ($M_{\rm gas}$), redshift ($z$), half-SFR radius, half-stellar-mass radius, and half-dust-mass radius. \\
\indent We first tuned an ANN using all ten features (``full'' model). We then trained ten further nine-feature ANNs, using the same architecture but with a different feature removed. In all cases, all input features were normalised, as described in Equation \ref{eq:feature_norm}. We compared the loss values of each of the nine-feature ANN to that of the ``full'' model. We eliminated features that did not significantly impact the loss values of the model.
It is reassuring that the features that were
determined to significantly affect the model
are all properties that should matter according
to physical intuition (e.g. SFR, dust mass).
Redshift is not important because we are predicting
SEDs in the rest frame.
\\
\indent In Table \ref{tab:corrmatrixTNG}, we show the correlation matrix of the galaxy properties used in \anngelina. Correlations are calculated as the covariance of each variable pair divided by the product of their standard deviations. A correlation with an absolute magnitude close to 1.0 indicates very strong correlation, while an absolute magnitude closer to 0.0 indicates a weak correlation. While some properties are strongly correlated
(most notably stellar and dust masses), \anngelina's performance is impacted negatively if any of the listed properties are omitted from the input feature set,
indicating that even strongly correlated features contribute non-redundant information.

\section{Results and discussion} \label{sec:results}
\subsection{Characterising \anngelina's overall performance}
We assess the performance of \anngelina using a test set of TNG50 galaxies that were not included in the training set. 
Here, we define the metrics used to characterise how well SEDs are predicted. We define the offset of the predicted (log-space) SED from the target (\skirt-predicted) SED at a given wavelength $\lambda$ as follows:
\begin{equation}\label{eq:offset_def}
\rm{Offset_{\lambda}/dex} = \log_{\rm 10} (E_{\lambda, \mathrm{predicted}}/E_{\lambda, \mathrm{true}}).
\end{equation}

\noindent For each galaxy in the test set, we take the median absolute offset value across all $200$ wavelengths, and call this the MAE:

\begin{equation}\label{eq:mae_def}
\mathrm{MAE/dex} = \mathrm{median}
\left| \rm{Offset_{\lambda}/dex} \right|.
\end{equation}

\noindent We use the MAE as a single value characterising the SED prediction for each galaxy. We note that, because of our definition, the MAE is essentially equivalent to the Median Absolute Percentage Error in the low-error regime.

\indent In Figure \ref{fig:nnfits}, we show several examples of target and predicted SEDs for galaxies in our test set. From left-to-right, we show predicted SEDs with MSEs at the $\approx$ $2^{\rm nd}$, $50^{\rm th}$, and $98^{\rm th}$ percentiles of the MSE distribution, where the $98^{\rm th}$ percentile refers to the worst of the three predictions, or the highest MSE. The top panels show the true SED calculated using \skirt in cyan and that predicted by \anngelina in black. The bottom panels show the logarithm of the prediction error
versus wavelength. The MSE and MAE are quoted at the tops of the columns. For the $2^{\rm nd}$ percentile example, the true and predicted SEDs are essentially indistinguishable -- the SED is predicted to within $\sim0.05\,\rm{dex}$ ($\sim12\%$) across the full UV-mm. The MAE is $0.0109\,\rm{dex}$ ($2.5\%$). For the $50^{\rm th}$ percentile example, \anngelina overpredicts the SED at both UV and FIR wavelengths, but the error is at most $\sim0.15\,\rm{dex}$ ($\sim41\%$). The MAE is $0.0643\,\rm{dex}$ ($16\%$). The $98^{\rm th}$ percentile example displays a catastrophic failure. The UV luminosity density is underpredicted by an order of magnitude. The MAE is $0.1638\,\rm{dex}$ ($46\%$).

\begin{figure}
\centering
\includegraphics[trim={0.1cm 0.2cm 0.1cm 0.2cm},clip,width=\linewidth]{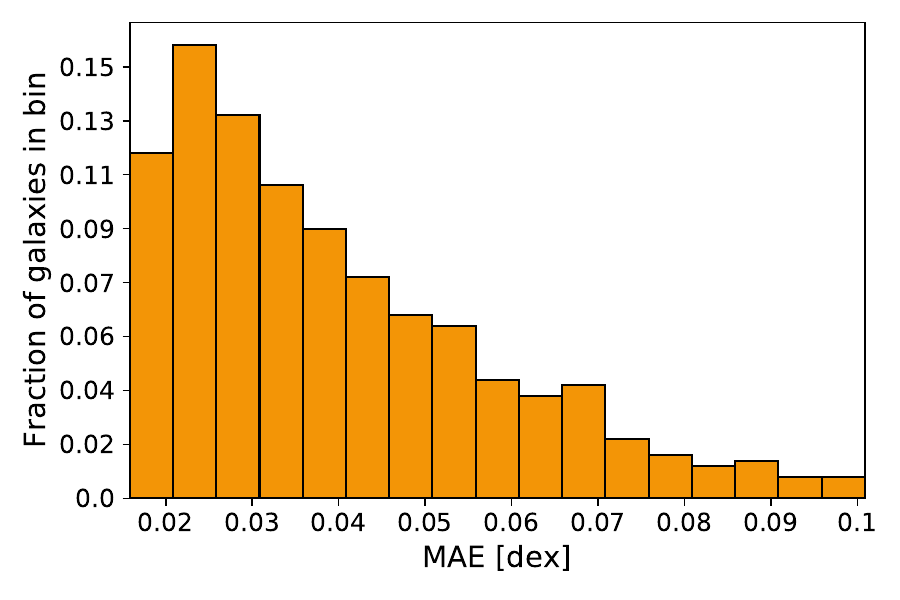}
\vspace{-0.5cm}
\caption{Distribution of the median absolute error, calculated for each galaxy using Equation \ref{eq:mae_def}. 
The top 2 per cent and bottom 2 per cent of values are excluded for clarity. On average, \anngelina predicts the irradiance to within 0.06 dex, but there is a large spread from galaxy to galaxy.}
\label{fig:loggedmaehist}
\vspace{-0.81cm}
\end{figure}

\begin{figure*}
\centering
\includegraphics[trim={0 0.4cm 0 0}, clip, width=\linewidth]{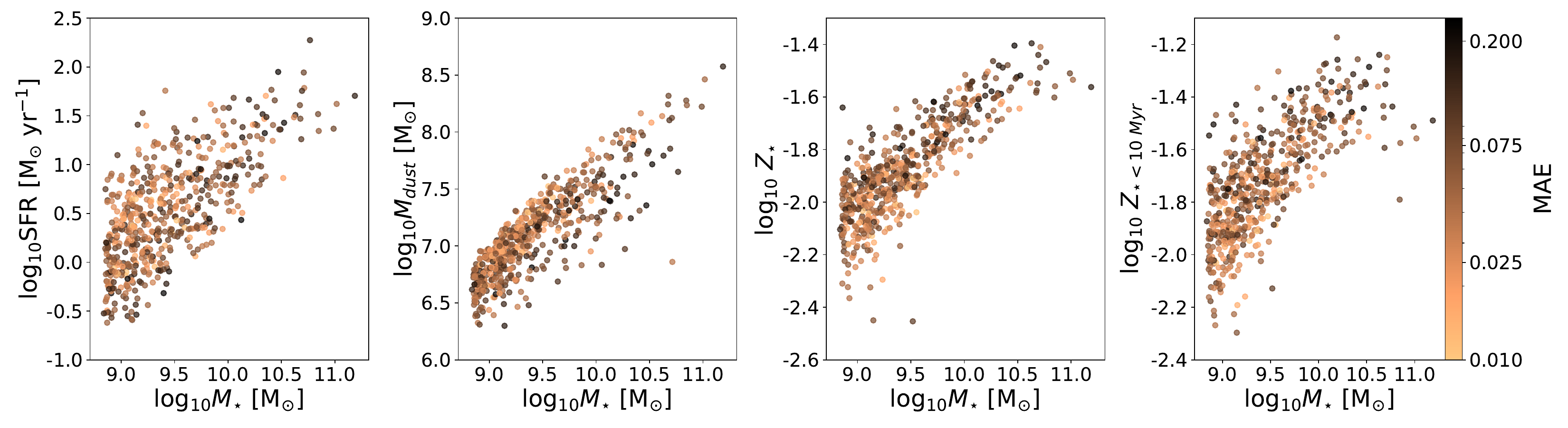}
\vspace{-0.2cm} 
\caption{TNG50 galaxies in the test set, color-coded by MAE value. The uniformity of MSE values in all feature-feature planes indicates that \anngelina performs uniformly well across the parameter space probed.}
\label{fig:mse_plane}
\end{figure*}

We compile MAE values for all test-set galaxies in Figure \ref{fig:loggedmaehist}. \anngelina performs well on the TNG50 dataset we employ, with an average MAE of $\sim0.06\,\rm{dex}$ ($15\,\%$), with the worst predictions having an MAE of $\sim0.2\,\rm{dex}$ ($60\,\%$).
In Figure \ref{fig:mse_plane}, we show 2D projections of the input feature space colored by MSE. If a particular region of parameter space were problematic for \anngelina, there would be local maxima in the MSE distribution. We see no such local maxima. This indicates that \anngelina does not exhibit higher MSE values in any one part of the parameter space; rather, it predicts the SEDs uniformly well throughout the parameter space probed. The scatter in the predictions may be due to additional variables not included in the analysis.

\begin{figure*}
\centering
\includegraphics[width=\linewidth]{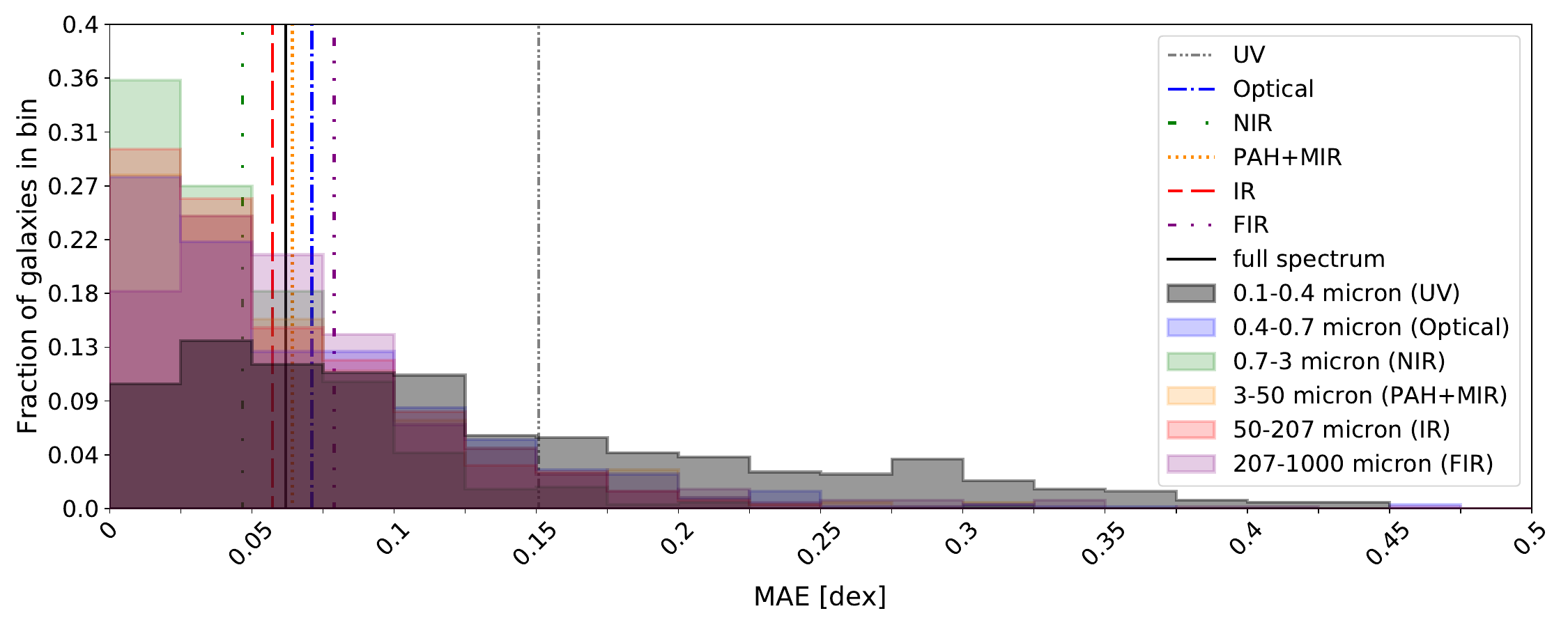}
\caption{Distributions of the median absolute error between the true and predicted SED per wavelength band for each galaxy in the test set. The vertical lines indicate the average MAE within each wavelength bin. The UV (\emph{grey dash-dot-dotted line}) has a significantly higher average MAE than all the other wavelength bands. This is due to the UV being most highly attenuated by dust, as dust attenuation is anisotropic. The other wavebands exhibit MAE values of $\sim 0.05-0.07$ dex.
The solid black line indicates the average MAE across the full wavelength spectrum.}
\label{fig:maebands}
\end{figure*}

\subsection{\anngelina's performance in different wavelength regimes}
As seen in Figure \ref{fig:nnfits}, for a given SED, \anngelina's accuracy varies with wavelength. In the examples shown, \anngelina appears to struggle to reproduce rest-frame UV emission more than other wavelengths. In this section, we characterise distributions of MAE values within individual wavelength ranges for the whole test sample. \\
\indent We divide SEDs into the following wavelength ranges: $0.1-0.4\,\micron$ (UV), $0.4-0.7\,\micron$ (optical), $0.7-3\,\micron$ (near-IR; NIR), $3-50\,\micron$ (mid-IR; MIR), $50-207\,\micron$ (far-IR; FIR) and $207-1000\,\micron$ (sub-mm). For each galaxy in the test set, we calculate the MAE within each of these wavelength ranges, following Equation \ref{eq:mae_def}. In Figure \ref{fig:maebands}, we compile these values (note that the worst 4\% of values have been omitted for clarity). Vertical lines indicate the average MAE across test galaxies within each of these wavelength bands. The UV stands out from the others, with a significantly higher average MAE of $0.15\,\rm{dex}$ ($41\%$), while all other wavelength bands hover around $0.06\,\rm{dex}$ ($15\%$). This is expected, as the UV emission depends much more strongly on the geometry of the observed simulated galaxy since the UV is highly attenuated by dust grains, and the amount of attenuation can vary considerably with the line of sight.

\subsection{Limitations of our approach}\label{sec:limitations}
There are a few considerations to highlight to potential users of \anngelina. First, all SEDs used to train \anngelina are for ``face-on'' projections of TNG50 galaxies (i.e.\,along the angular momentum axis). 
Consequently, for galaxies with disc-like morphologies, our dataset is biased toward the least-obscured lines-of-sight, and heavily obscured lines-of-sight (e.g.\,edge-on discs) are thus under-represented.
The SEDs of simulated galaxies with exactly the same values for the input features can vary due to viewing-angle effects: dust attenuation is non-isotropic, especially in the UV. For this reason, if the ``raw'' predictions from \anngelina are used, the scatter in the predicted SEDs will be unrealistically low.
One cannot simply perturb the predicted photometry by adding random noise to account for this effect, as the prediction errors for flux values in different bands should be correlated (e.g.\,for an edge-on galaxy, \anngelina is likely to systematically overpredict the UV-optical flux).
In future work, we will employ a larger training set with multiple lines-of-sight and incorporate some geometric features to attempt to account for viewing angle effects in
the ANN's predictions. Since there is intrinsic scatter to be expected when predicting an SED given a set of galaxy properties, we will also attempt to use normalizing flows to sample the distribution instead of providing the model a single value as we have done here.\\
\indent Second, \anngelina was trained on TNG50 galaxies sampling a restricted region of parameter space (simulated galaxies with $M_{\star}>10^{8.7}\,\rm{M_{\odot}}$ on or above the main sequence), and it should only be applied within this region. Applying \anngelina to simulated galaxies outside of this parameter space may result in high prediction error, even when these galaxies are simulated with exactly the same code (i.e.\,numerical method and sub-grid models) and resolution.
\indent Moreover, only a single simulation -- and thus single code and single resolution -- was used. It is possible that the prediction error would be significantly greater
if \anngelina were applied to simulations that employ the \emph{IllustrisTNG} model but different resolution or/and simulations that employ a different numerical method or/and
sub-grid models. In this proof-of-concept work, we have opted to employ only a single simulation, but in future work, we will explore how well we can cross-apply
our emulator to other simulations. \\
\indent Finally, we note that various assumptions are `baked in' to the MCRT calculations and thus \anngelina. For example, it is necessary to make assumptions about the stellar populations (stellar initial mass function and single-age stellar population SED templates) and dust (optical properties, dust-to-metal ratio, temperature above which dust is destroyed, and sub-grid dust distribution). These assumptions can affect the resulting SEDs, and depending on the science question of interest, they may affect the quantitative or even qualitative results. We have explored the sensitivity of our results to such assumptions in various previous works \citep[e.g.][]{hayw11,Snyder2013,Safarzadeh2017irx}. Exploring the impact of such assumptions is beyond the scope of this proof-of-concept work. Instead, as noted in the introduction, we have addressed the following question: ``given a set of 3D MCRT calculations with fixed assumptions (regarding e.g.\,the stellar initial mass function, single-age stellar population SED templates, and dust model) performed on galaxies selected from a single cosmological simulation, how well can an ANN emulate the MCRT calculations to predict integrated UV-mm SEDs of the simulated galaxies?'' \\
\indent It is important that users understand that the results will in general depend on the assumptions used in the MCRT calculations to generate the training set. For this reason, the reported errors \emph{do not} include systematic errors associated with the MCRT assumptions and thus underestimate the `true' error. These errors only represent how imperfectly the ANN can emulate MCRT calculations for this single simulation and set of MCRT assumptions. If, for example, one desires to predict SEDs assuming SMC-like dust, it is necessary to generate a training set using SMC-like dust in the MCRT calculations and then train a new ANN rather than using the ANN that we have made publicly available.

\section{Conclusions}\label{sec:conclusions}
In this work, we have presented an ANN-based emulator to predict UV-mm SEDs of simulated galaxies. We trained the ANN on a sample of SEDs
generated in previous work by performing dust MCRT on galaxies from the TNG50 simulation. We find that the ANN performs well
at predicting the SEDs of simulated galaxies in the test set --
the average MAE is 0.06 dex.
For the vast majority of simulated galaxies in the test set, \anngelina can predict the flux across the full UV-mm wavelength range with
a maximum relative error of $\la 50$ per cent. The UV dominates the prediction error, likely because the viewing angle dependence
is strongest in this wavelength regime, and our input feature set includes no information about viewing angle.
Our results demonstrate that our ANN-based emulator is a promising computationally inexpensive alternative to
performing MCRT in order to predict UV-mm SEDs of simulated galaxies.

\section*{Acknowledgements}
The authors thank the anonymous referee for their useful comments, which helped improve the clarity of the manuscript. SS is supported by the NASA FINESST fellowship award 80NSSC22K1589. SS acknowledges support from the CCA Pre-doctoral Program, during which this work was initiated. The Flatiron Institute is supported by the Simons Foundation. JHW acknowledges support from NASA grants NNX17AG23G, 80NSSC20K0520, and 80NSSC21K1053 and NSF grants OAC-1835213 and AST-2108020.

\section*{Data Availability}\label{sec:dataavail}

A github repository containing the tools to run our RT emulator is available at
\href{https://github.com/snigdaa/runANNgelina}{https://github.com/snigdaa/runANNgelina}.
Our cleaned data are available at the github repository linked above.



\bibliographystyle{mnras}
\bibliography{main}

\bsp 
\label{lastpage}
\end{document}